\documentclass[twocolumn,showpacs,prb,aps,amssymb]{revtex4}

\usepackage{graphicx}
\usepackage{bm}

\newcommand{\dfrac}[2]{\frac{\strut \displaystyle{#1}}{\displaystyle{#2}}}

\begin{document}

\title{Quantum Melting of Incommensurate Domain Walls in Two
Dimensions}

\author{Tsutomu Momoi}
\affiliation{Institute of Physics, University of Tsukuba, Tsukuba,
Ibaraki 305-8571, Japan}
\date{February 28, 2002}

\begin{abstract}
Quantum fluctuations of periodic domain-wall arrays in
two-dimensional incommensurate states at zero temperature are
investigated using the elastic theory in the vicinity of the
commensurate-incommensurate transition point. Both stripe and
honeycomb structures of domain walls with short-range interactions
are considered. It is revealed that the stripes melt and become a
stripe liquid in a large-wall-spacing (low-density) region due to
dislocations created by quantum fluctuations. This quantum melting
transition is of second order and characterized by the
three-dimensional XY universality class. Zero-point energies of
the stripe and honeycomb structures are calculated. As a
consequence of these results, phase diagrams of the domain-wall
solid and liquid phases in adsorbed atoms on graphite are
discussed for various domain-wall masses. Quantum melting of
stripes in the presence of long-range interactions that fall off
as power laws is also studied. These results are applied to
incommensurate domain walls in two-dimensional adsorbed atoms on
substrates and in doped antiferromagnets, e.g. cuprates and
nickelates.
\end{abstract}

\maketitle

\section{Introduction}

It has been established that a macroscopic number of domain walls
(``discommensurations'' or ``soliton excitations'') appear in
incommensurate states and form a
global soliton-lattice structure.\cite{Bak,PokrovskyT,%
BakMVW,Villain,Coppersmith,FisherF,Kardar,Halpin-Healy}
Two-dimensional (2D) solids of adsorbed atoms on graphite have
been studied as a typical example of commensurate-incommensurate
phase transitions (C-IC
transitions).\cite{Birgeneau,HongPMBHS,Wiechert,Greywall} As 2D
incommensurate soliton lattices, theoretical
works\cite{PokrovskyT,BakMVW} proposed stripe and honeycomb
structures, and both of them were observed in
experiments.\cite{Birgeneau,Wiechert,Grimm} Since domain walls are
made from excess atoms (for heavy walls), the domain-wall mass can
be controlled in experiments by changing atoms from Xe to $^3$He
(or H$_2$) and, thereby, quantum fluctuation can be increased.
Recently, striped anti-phase domain walls were observed in doped
nickel oxides and cupper oxides,\cite{Tranquada,Yamada} in which
the stripe structure shows incommensurate short-range order. The
effect of quantum fluctuations on the 2D domain-wall structures
was hence attracting attention, but in theoretical understanding,
little is known about quantum disordered striped or honeycomb
states. Striped domain walls were also found in a frustrated 2D
quantum spin system that shows magnetization
plateaux.\cite{MomoiT}

Neutron scattering measurements in the striped incommensurate
phases showed that incommensurability is proportional to the
excess density $\delta n$ in adsorbed atoms\cite{Wiechert} and to
Sr concentration $x$ in La$_{2-x}$Sr$_x$CuO$_4$ for the underdoped
region, $x<1/8$.\cite{Yamada,Fujita} The nature of one domain wall
is, hence, unchanged for the whole density region, and the
domain-wall spacing only changes depending on the excess (or hole)
density. It was argued for doped antiferromagnets that one domain
wall behaves like a quantum elastic
string.\cite{NayakW,EskesOGSZ,KivelsonFE}
In adsorbed atom systems, domain walls are not pinned by the
substrate periodic potential, but floating, and hence we can also
adopt this string model to describe the single domain wall.

The stability of domain-wall arrays is controlled by effective
interactions between walls and fluctuations among
them\cite{Villain,Coppersmith}. When the wall-wall repulsive
interactions are strong enough, the domain walls form a regular
array and thereby becomes a solid with incommensurate long-range
order (LRO). In experiments, this LRO is observed as sharp
incommensurate Bragg peaks, and in the elastic theory for domain
walls\cite{Villain,Coppersmith} this LRO is characterized by
finite stiffness. On the other hand, when fluctuations are strong,
dislocations proliferate and thereby make the domain-wall ordering
short ranged. In this case, the domain walls are irregular and
show a ``liquid''-like behavior with exponential decay of
correlations, in which incommensurate scattering peaks have finite
width. In the elastic theory, this short-range domain-wall order
is characterized by vanishing of stiffness. It was shown that
thermal fluctuations induce a phase transition from the striped or
honeycomb (ordered) solid to a short-range-ordered liquid at
finite temperature and the stripe liquid phase remains until zero
temperature at the onset of the C-IC
transition.\cite{Villain,Coppersmith,Kardar} Quantum fluctuation
was not taken into account previously with regard to melting of
domain-wall structures.

In the classical theory for domain walls in incommensurate
adsorbed atoms, the wall-wall repulsive interaction and the wall
intersection energy determine the ground-state phase diagram of
domain-wall structures.\cite{BakMVW} In the stripe structure,
walls with spacing $l$ repel each other by the exponential
interaction\cite{Villain} $\alpha \kappa \exp(-\kappa l )$. Here
$\kappa$ denotes the inverse of the domain-wall width and
$\alpha=c_1 Qa^2 Y$, where $c_1$ denotes a constant of order
unity, $Q$ the degeneracy of commensurate domains, $a$ the unit
length of atoms in commensurate domains, and $Y$ the microscopic
Young's modulus of the adsorbate. At zero temperature, the striped
domain walls form a regular parallel array with the energy per
unit area
\begin{equation}
E_{\rm s}=\dfrac{\zeta_0-\zeta}{l} +\dfrac{\alpha\kappa}{l} \exp
(-\kappa l),
\end{equation}
where $\zeta_0$ and $\zeta$ denote the energy and the chemical
potential of one wall per unit length. On the other hand, the
honeycomb structure has the wall-intersection energy\cite{BakMVW}
$f_{\rm I}$ per intersection. The total energy of a regular
honeycomb domain walls per unit area is
\begin{equation}
E_{\rm h}=\dfrac{2(\zeta_0 - \zeta)}{\sqrt{3}l}
 + \dfrac{4 f_{\rm I}}{3\sqrt{3} l^2}
 + \dfrac{\sqrt{3} \alpha \kappa}{l} \exp (-\sqrt{3} \kappa l),
\end{equation}
where $l$ is the length of one side of the unit hexagon. For each
structure, the length $l$ is determined by minimizing the energy
with fixed $\zeta$. The ground-state structure is selected by
comparing the energies with determined $l$'s. When the
intersection energy $f_{\rm I}$ is positive, the stripe array is
more stable than the honeycomb structure for large wall-spacing
$l$ due to absence of the intersection energy, whereas the
honeycomb structure is favored for moderate $l$ due to rapid decay
of the wall-wall interaction. If the intersection energy is
negative, only honeycomb states appear in the whole density
region.\cite{BakMVW}

\begin{figure}[tbp]
\includegraphics[width=2.5in]{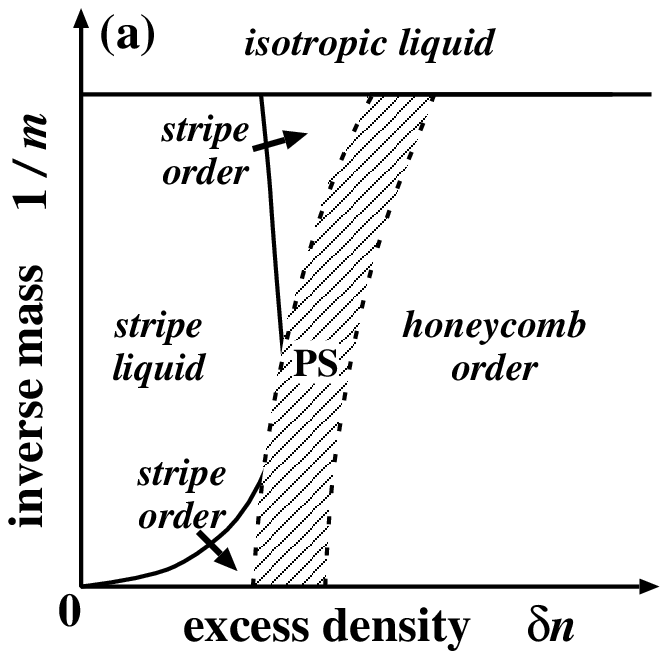}
\includegraphics[width=2.5in]{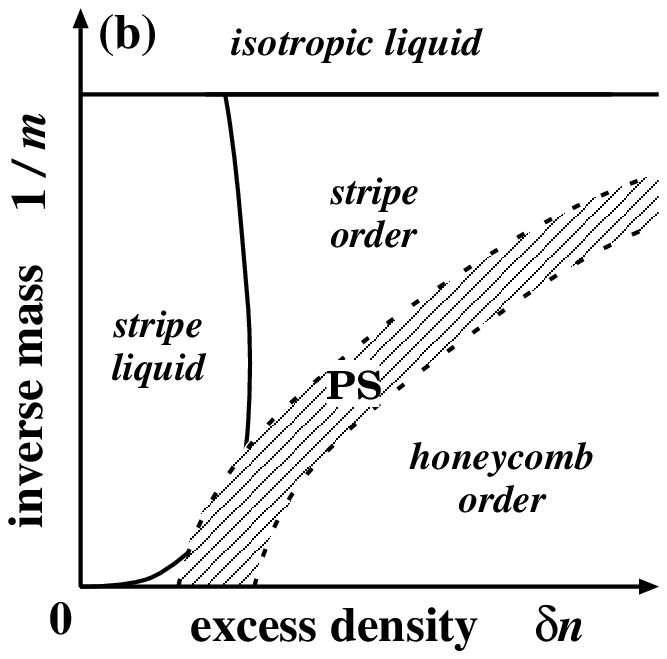}
\caption{\label{fig:pd_final}Schematic phase diagrams of
incommensurate phases in 2D adsorbed atoms on substrates. The
wall-intersection energy is set to be positive but not very large
and the domain-wall width $\kappa^{-1}$ (a) fixed and (b)
$m$-dependent. (See text in Sec.\ \ref{sec:PD}.) The phase
transition between the stripe liquid phase and the stripe ordered
one is of second order. In the shaded regions, the phase
separation (PS) occurs between the stripe phases and the honeycomb
one.}
\end{figure}

In this paper, we study quantum fluctuations in domain-wall
structures taking account of spontaneous creation of dislocations
at zero temperature.
We use an elastic theory to discuss striped domain walls or
honeycomb ones. Considering the effect of dislocations in the 2+1D
space, we discuss a quantum-melting transition of the stripe
structure, which occurs as a consequence of the proliferation of
dislocations. It is shown that, for a large-wall-spacing region,
the striped structure with short-range wall-wall interactions is
unstable against quantum fluctuations and becomes a stripe liquid
even at zero temperature. This melting transition is a continuous
one in the universality class of the three-dimensional (3D) XY
model. We point out that this quantum disordered stripe phase was
presumably observed in neutron scattering experiments of adsorbed
H$_2$ or D$_2$ monolayer on graphite,\cite{Freimuth} and in those
of doped cupper oxides and nickel oxides.\cite{Yamada,Fujita} For
2D adsorbed atoms, our results indicate that when the wall
intersection energy $f_{\rm I}$ is positive, the stripe-liquid
phase appears between the commensurate and stripe-ordered phases.
The zero-point energies are also estimated for both stripe and
honeycomb structures, and finally, possible phase diagrams for
incommensurate 2D atoms on surfaces are given for various strength
of quantum fluctuations. For convenience of readers, we show the
phase diagrams obtained in this paper in Fig.\ \ref{fig:pd_final}
in advance. The inverse of the domain-wall mass $1/m$ and the
excess density $\delta n$ are taken as variables. The parameter
$1/m$ represents the strength of quantum fluctuations and the
limit $1/m\rightarrow 0$ corresponds to the classical limit. A
phase separation occurs in a certain density region between the
stripe phases and the honeycomb one. On the other hand, if stripes
have long-range wall-wall interactions that fall off as power laws
$r^{-q+1}$ with $q<5$, the stripe structure is stable in low
density and unstable in high density. This tendency to ordering
is, thus, opposite to the case of short-range interaction systems.

We discuss the stripe structure with short-range interactions in
section \ref{sec:stripe} and the honeycomb structure in section
\ref{sec:honeycomb}. Phase diagrams for adsorbed atoms on
substrates are proposed in section \ref{sec:PD}. In section
\ref{sec:power}, we discuss stripes with long-range interactions
that fall off as power laws. Section \ref{sec:discuss} contains
discussions.

\section{Striped domain walls}\label{sec:stripe}
In this section, we discuss quantum fluctuation of stripes and
show how a quantum stripe liquid appears.

\subsection{Hamiltonian of quantum stripes with dislocations} Let
us consider first the long-wavelength fluctuations of stripes
which do not contain any dislocation. For simplicity, we consider
striped domain walls on a rectangular substrate, which are aligned
along the $y$-axis. It was argued for doped
antiferromagnets\cite{NayakW,EskesOGSZ,KivelsonFE} that a single
domain wall behaves as a quantum elastic string with the wall mass
$m$ per unit length and the interfacial stiffness $\gamma$.
In adsorbed atom systems, domain walls are floating and hence this
string model can be also adopted. Here the mass $m$ relates to the
mass of adsorbed atoms, and the stiffness $\gamma$ comes from both
the line tension of the walls and the anisotropy of substrates. In
a stripe pattern, domain walls interact with each other by both
the exponential repulsion and hard-core potential, and hence
stiffness $K_x$ appears between walls. In the continuum limit of
the displacement field, the Hamiltonian for stripes without any
dislocation has the form\cite{Zaanen}
\begin{equation}\label{eq:H}
H_0=\frac{1}{2}\int d^2 r \left\{ \dfrac{lp^2}{m} +
\dfrac{\gamma}{l} (\partial_y u)^2 + K_x ( \partial_x u )^2
\right\}
\end{equation}
up to quadratic terms, where $u$ denotes the $x$ displacement of
the wall from the straight position and $p$ is the conjugate
momentum. A path integral representation leads to $Z_0=\int {\cal
D}u(\tau,{\bf r}) \exp(-S_0)$ with the effective action
\begin{equation}\label{eq:effect1}
S_0[u]=\frac{1}{2\hbar}\int d\tau d^2r \sum_{\alpha=\tau,x,y}
K_\alpha (
\partial_\alpha u )^2,
\end{equation}
where $K_\tau=m/l$ and $K_y=\gamma/l$. The functional integral has
ultra-violet (UV) momentum cutoffs $|q_\tau|\le \pi c/a$,
$|q_x|\le \pi /Ql$, and $|q_y|\le \pi /a$, where
$c\equiv\sqrt{\gamma/m}$, and $Q$ denotes the degeneracy of
commensurate ground states. With rescaling $T=c\tau$,
$X=(\gamma/lK_x)^{1/2}x$, and $Y=y$, we have the isotropic
effective action
\begin{equation}\label{eq:effect2}
S_0[u]=\dfrac{K}{2\hbar} \int dT d^2R
\sum_{\alpha=\tau,x,y}(\partial_\alpha u)^2
\end{equation}
with $K=(m K_x/l)^{1/2}$.

\begin{figure}[tbp]
\includegraphics[width=6.9cm]{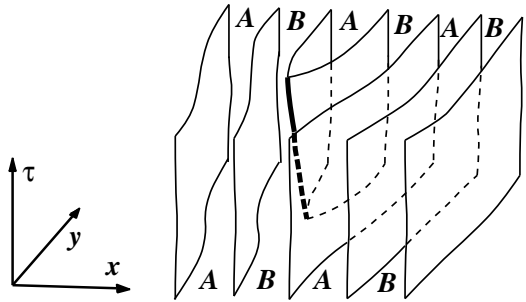}
\caption{\label{fig:dislocation}Dislocation line of stripes in the
2+1D space when the commensurate domains are doubly degenerate
($Q=2$). Two types of commensurate domains are labelled
 with $A$ and $B$.}
\end{figure}
Equation (\ref{eq:H}) is not the full Hamiltonian, since real
stripes have dislocations (defects) even at zero temperature owing
to quantum fluctuations. Hence we next take into account the
singular part in $u$ that comes from dislocations. The
dislocations become lines (or loops) in the 2+1D space as shown in
Fig.\ \ref{fig:dislocation}. $Q$ domain walls merge together into
a dislocation so that each dislocation is consistent with the
domain structure. The strength of dislocation lines can be
observed\cite{NelsonT,ChaikinL} by a loop integral over derivative
of the displacement field on a closed loop $\Gamma$
\begin{equation}
\oint_\Gamma du =
Qls, \label{burger}
\end{equation}
where $s~(=0$, $\pm 1$, $\pm 2,\cdots)$ denotes the number of
dislocation lines enclosed by the loop $\Gamma$. The field $u$ can
be divided into the smooth part $u_{\rm smooth}$ and the singular
part $u_{\rm sing}$ that describes dislocation lines. Thus Fourier
transform of the gradient of $u$ can be written as
\begin{equation}\label{eq:field}
i{\bf q}u_{\rm smooth}({\bf q})+i Ql {\bf
q}\times {\bf n}({\bf q})/q^2
\end{equation}
in the continuum limit, where ${\bf q}=(q_x,q_y,q_\tau)$ and ${\bf
n}({\bf q})$ denotes Fourier transform of dislocation density
vectors. Inserting eq.\ (\ref{eq:field}) into the gradient vector
$\nabla u$ in eq.\ (\ref{eq:effect2}), we obtain the total
effective action\cite{NelsonT}
\begin{equation}
S=S_0[u_{\rm smooth}] + S_{\rm D}[{\bf n}]
\end{equation}
with
\begin{equation}
S_{\rm D}[{\bf n}]=\dfrac{1}{2\hbar}\int \dfrac{d^3 q}{(2\pi)^3}
\dfrac{(Ql)^2 K}{q^2} |{\bf n({\bf q})}|^2.
\end{equation}

\subsection{Criterion for quantum melting}\label{sec:stripe_melting}
After transforming the displacement field as $\tilde{u}=2\pi
u/Ql$, the system has periodicity $2\pi$. By rescaling
$\tilde{\tau}=c\tau$, $\tilde{x}=ax/Ql$ and $\tilde{y}=y$, the
effective action is
\begin{equation}\label{rescaled-action}
S_0=\dfrac{1}{2\hbar}\int d\tilde{\tau}d^2 \tilde{r} \sum_\alpha
\tilde{K}_\alpha (\tilde{\partial}_\alpha \tilde{u})^2
\end{equation}
with the couplings
\begin{eqnarray}
\tilde{K}_\tau &=& \tilde{K}_y = Q (Ql)^2 \sqrt{m\gamma}/(2\pi)^2
a,
\nonumber\\
\tilde{K}_x &=& (Ql) K_x/(2\pi)^2 c a, \label{rescaled-coeff}
\end{eqnarray}
where cutoffs are set as $\Delta \tilde{\tau}=\Delta
\tilde{x}=\Delta \tilde{y}=a$. In these notations, dislocations
are quantized as
\begin{equation}
\oint_\Gamma d\tilde{u} = 2\pi s. \label{burger2}
\end{equation}
The final forms (\ref{rescaled-action}) and (\ref{burger2}) are
equivalent to a spatially anisotropic 3D XY model in the vortex
loop representation.\cite{NelsonT,Nelson} Regarding the
displacement field as the phase $\theta(r) = 2\pi u(r)/Ql$, we
thus mapped dislocations of stripes onto vortices of phases, and
the 2D quantum stripe system onto the anisotropic 3D XY model with
couplings $2 a \tilde{K}_\tau$, $2 a \tilde{K}_x$ and $2 a
\tilde{K}_y$.

In the absence of dislocation lines, the Hamiltonian
(\ref{eq:effect1}) has only quadratic terms, and consequently, the
stripes have finite stiffness at any value of parameters $l$ and
$m$. Zaanen\cite{Zaanen} argued that dislocations cannot
proliferate in quantum stripes at zero temperature, but we
disagree with his points. His argument about dislocations is based
on a classical picture of Kosterlitz-Thouless
transition\cite{KosterlitzT} and can be applied only to classical
striped systems. Since the quantum system is mapped onto a 2+1D
one at finite temperature, gain of quantum entropy can overcome
loss of the core energy of dislocations in a specific parameter
region. In 3D superfluid, which is equivalent to the 3D XY model,
Feynman\cite{Feynman} proposed that vortex loops are responsible
for transition. Then it was discussed that dislocation loops
reduce superfluidity to zero at the critical
temperature.\cite{NelsonT,Nelson} This picture was confirmed by
Monte Carlo simulations of the 3D XY model\cite{KohringSW} and the
critical exponent $\nu=0.67$ was correctly derived from a
renormalization-group treatment of vortex loops.\cite{Shenoy}

For the mapped 3D XY model, we know that there are ordered and
disordered phases in the parameter space of $\tilde{K}_\tau$,
$\tilde{K}_y$, and $\tilde{K}_\tau$. It is natural to expect that
the critical point is given by the relation
\begin{equation}\label{eq:criterion}
(\tilde{K}_x \tilde{K}_y \tilde{K}_\tau)^{1/3} \hbar^{-1} = C
\end{equation}
with a constant $C$ of order unity, and long-range order exists
for $(\tilde{K}_x \tilde{K}_y \tilde{K}_\tau)^{1/3} \hbar^{-1} >
C$. Due to the presence of dislocation lines, the stiffness of
stripes is reduced and the quantum disordered phase appears in
$(\tilde{K}_x \tilde{K}_y \tilde{K}_\tau)^{1/3} \hbar^{-1} < C$.
This melting is a continuous transition in the same universality
class as the 3D XY model. In the stripe-liquid phase, the
correlation length shows the form\cite{Guillou}
\begin{equation}
\xi \sim (n_{\rm c}-n)^{-\nu}
\end{equation}
with $\nu\approx 0.67$ near the critical density $n_{\rm c}$.
Though there are $Q$ degenerate ground states for each stripe
configuration, they can be changed from one to others by
repetition of transformation $u\rightarrow u+a$ and hence may not
affect properties of the phase transition.

\subsection{Phase diagram} To proceed with calculation of the
phase diagram, we need parameter dependence of $K_x$. Here, we
derive the form of $K_x$ with various approximations and thereby
obtain the phase diagram of the stripe liquid and ordered phases.

In the large-$m$ limit, the wall fluctuation is negligible and the
exponential repulsion dominates the stiffness
\begin{equation}
\label{exponential} K_x = \alpha \kappa^3 l \exp(-\kappa l).
\end{equation}
When the mass $m$ is finite, but large enough, the wall
fluctuation slightly modifies the effective repulsion form. Since
the quantum elastic string is equivalent to a 3D classical
Gaussian surface in the path-integral picture, we can use results
derived for the 3D Gaussian surface. From the
renormalization-group (RG) arguments\cite{BrezinHL,Sornette} in a
mean-field approximation,\cite{MF} the stiffness is derived in the
form
\begin{equation}
\label{exponential2} K_x= \alpha \kappa^3 l
\exp\left(-\frac{2\kappa l}{2+\omega}\right)
\end{equation}
for $\omega\leq 2$ with the dimensionless parameter
\begin{equation}\label{quantum_parameter}
\omega\equiv\frac{\hbar \kappa^2}{4\pi \sqrt{m\gamma}},
\end{equation}
which represents strength of quantum effects.

On the other hand, if the quantum wall-fluctuation is large, i.e.
$\omega\gg 1$, the hard-core repulsion becomes important,
producing {\it steric} force between walls.\cite{FisherF} For the
mean-field argument\cite{MF}, in which one surface is confined in
a limited space between perfectly plane two walls with distance
$2l$, a rigorous proof shows that the effective entropic repulsion
decays exponentially.\cite{BF} From rescaling $u'=u/l$ and
$\tau'=c \tau$, it comes out that the dimensionless parameter of
surface configuration is $\hbar /(l^2 \sqrt{m\gamma})$. We, hence,
conclude that $K_x$ depends on $l$ for large $l$ in the form
\begin{equation}\label{hard-middle}
K_x \simeq \exp\{-c_2 (m \gamma)^{1/4} l /\sqrt{\hbar} \},
\end{equation}
where $c_2$ is a positive constant. This is consistent with the
form derived from RG arguments\cite{Sornette} (see also Appendix
A)
\begin{equation}\label{hard-middle2}
K_x \simeq \Biggl(\dfrac{\kappa^2}{\omega}\Biggr)^{11/8}
l^{3/4}\exp\left( -\frac{\kappa l}{\sqrt{2\omega}} \right),
\end{equation}
which is valid for $\omega\geq 2$. Note that the above result is
derived in a kind of "mean-field" approximation.\cite{MF} Recently
Zaanen\cite{Zaanen} treated fluctuations of many walls in the
harmonic approximation and estimated the stiffness between walls
from Helflich's self-consistent condition.\cite{Helfrich} In the
large-$l$ limit, his result shows the stretched exponential form
$K_x \simeq \exp\{-c_3 (\sqrt{m \gamma}/\hbar)^{1/3}l^{2/3} \}$,
where $c_3$ is a positive constant. It might be unclear whether
his argument gives the exact result at the delicate dimensionality
($d=2$), where the Gaussian surface is in the upper critical
dimension $D~(=d+1)=3$. In the moderate distance $l$, however, the
self-consistent condition gives the same exponential form as eq.\
(\ref{hard-middle2}). Since the difference between the simple
exponential and stretched exponential forms appears only in an
extremely large-spacing region, we use the simple exponential one
in the following discussions.

The large $m$ and small $m$ regions in which the original
exponential repulsion and the hard-core one dominate,
respectively, can be defined from the behavior of the deviation
$\Delta u$ of the displacement. Using the mean-field
approximation\cite{MF} and applying the harmonic approximation to
the potential created by exponential repulsions between adjoining
walls, we have the Hamiltonian of a single wall
\begin{equation}
H=\dfrac{1}{2} \int dy \biggl\{ \dfrac{p^2}{m} + \gamma(\partial_y
u)^2 + M^2 u^2 \biggr\}
\end{equation}
with $M^2=2\alpha \kappa^3 \exp(-\kappa l)$.
The deviation of $u$ in the harmonic approximation is estimated as
\begin{equation}
[(\Delta u)^2]_{\rm h.a.} = \dfrac{\hbar c}{4\pi\gamma}
\log\biggl(\dfrac{\pi^2 \gamma}{a^2 M^2}\biggr) \approx
\frac{\omega l}{\kappa}
\end{equation}
for large $l$. On the other hand, if the quantum string wanders a
lot and repels from the adjoining straight lines by the hard-core
repulsion, the deviation is proved\cite{BF} to be $(\Delta u)^2
\sim l$. Since the deviation $(\Delta u)^2$ has length dimension
as $l^2$, it should be scaled as $(\Delta u)^2 =l^2 F(\hbar /l^2
\sqrt{m\gamma})$. To incorporate the $l$ dependence for large $l$,
we set $F(x)\sim \sqrt{x}$. The quantum wall with hard-core
repulsion, thus, has the deviation
\begin{equation}
[(\Delta u)^2]_{\rm h.c.} \approx \dfrac{\sqrt{\omega}l}{\kappa}.
\end{equation}
By comparison with $[(\Delta u)^2]_{\rm h.a.}$ and $[(\Delta
u)^2]_{\rm h.c.}$, it is clear that, for $\omega \ll 1$, the
deviation of the wall in the harmonic potential is small enough
and the hard-core repulsion does not work, i.e., the wall is
controlled by the exponential repulsion, while for $\omega \gg 1$
the hard-core repulsion controls the wall wandering and the steric
interaction works. This result is consistent with that from RG
treatments in the large $l$ limit,\cite{BrezinHL,Sornette} where
the relevant interaction changes from the exponential one
[eq.~(\ref{exponential2})] to the hard-core one
[eq.~(\ref{hard-middle2})] at $\omega=2$. This change of dominant
interactions is a crossover for moderate $l$.

\begin{figure}[tbp]
\includegraphics[width=2.5in]{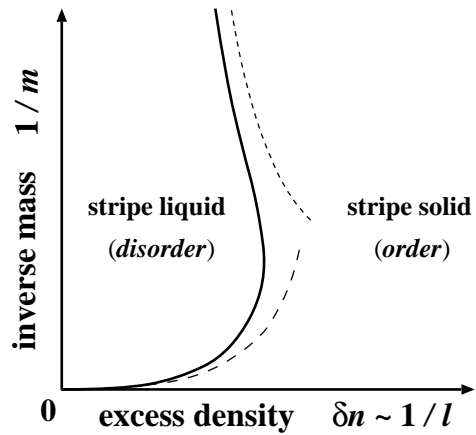}
\caption{\label{fig:pd_stripe}Possible phase diagram of stripes.
The dotted line and dashed one denote asymptotic forms of the
phase boundary for small $m$ [eq.\ (\ref{asym:small})] and large
$m$ [eq.\ (\ref{asym:large})], respectively.}
\end{figure}

Here we start discussion of the phase diagram. In sec.\
\ref{sec:stripe_melting}, we showed that stripes become
short-range ordered in $(\tilde{K}_x \tilde{K}_y
\tilde{K}_\tau)^{1/3} \hbar^{-1} < C$, where $C$ is a constant of
order unity. Inserting eqs.\ (\ref{rescaled-coeff}),
(\ref{exponential2}), and (\ref{hard-middle2}) into this
criterion, we conclude that the stripes melt for large $l(>l_c)$
and become a short-range ordered state, i.e., a stripe liquid.
(Though this equation is satisfied by two values of $l_{\rm c}$,
this equation is asymptotically correct in the large $l$ limit and
hence large $l_{\rm c}$ is the physically meaningful one.) We
expect that this conclusion does not rely on the approximation
method we employed. Even if the steric force decays in a stretched
exponential form, this statement still holds. The phase diagram of
the stripe solid and liquid phases is shown in Fig.\
\ref{fig:pd_stripe} for various mass $m$. In the large-mass limit,
which corresponds to the classical system, there is no liquid
phase. For large $m$, the critical length at the melting
transition is
\begin{equation}\label{asym:large}
l_c \simeq \dfrac{1}{\kappa}
\log \dfrac{Q^7\sqrt{\gamma}m^{3/2}}{a\hbar^3\kappa^5}
\end{equation}
and it rapidly shrinks with reducing the mass $m$. Due to a
crossover of interactions from the exponential one to the steric
one, the critical length turns to increase for small $m$,
satisfying
\begin{equation}\label{asym:small}
l_c \simeq \dfrac{\sqrt{\hbar}}{\sqrt{2\pi}(\gamma m)^{1/4}} \log
\dfrac{Q^7 \gamma^{3/4}}{a m^{1/4} \hbar^{3/2} \kappa^{33/16}}.
\end{equation}
The phase boundary, thus, shows reentrant behavior with changing
the parameter $1/m$.

From the above calculations, we also obtain the zero-point energy
of striped domain walls described by the Hamiltonian (\ref{eq:H}).
Using eqs.\ (\ref{exponential2}) and (\ref{hard-middle2}), the
zero-point energy of stripes per area is estimated as
follows\cite{Sornette} (see appendix A): For small $m$, i.e.,
$\omega \gg 1$,
\begin{equation}\label{eq:zero-e1}
\Delta E_{\rm s}\simeq\dfrac{\zeta_0}{l} + \dfrac{c_4
\sqrt{\gamma}}{a l^{5/4}} \Biggl(\dfrac{\hbar}{\sqrt{\gamma
m}}\Biggr)^{5/8} \exp\left\{ -\Biggl(\dfrac{2\pi \sqrt{\gamma
m}}{\hbar}\Biggr)^{1/2} l \right\},
\end{equation}
where $\zeta_0$ denotes the zero-point energy of one wall per
length given by $\zeta_0 = (\hbar \pi/4a^2) \sqrt{\gamma/m}$ and
$c_4$ is a positive constant, and for large $m$, i.e., $\omega \ll
1$,
\begin{equation}
\Delta E_{\rm s} \simeq \dfrac{\zeta_0}{l}
 + \dfrac{\hbar c_1 \kappa^2}{8\pi \sqrt{m \gamma}}
 \exp \left(-\frac{2\kappa l}{2+\omega} \right).
\end{equation}
In this approximation, the C-IC transition is of second order and
the chemical potential dependence of the excess density is $\delta
n \sim 1/l \sim [-\log (\zeta - \zeta_0)]^{-1} $ for any $m$. If
the stretched exponential form is adopted as the steric
interaction, $\delta n \sim [-\log (\zeta - \zeta_0)]^{-3/2}$ for
small $m$. This critical behavior may be modified if dislocations
of stripes are taken into account. In an exactly solvable 2D
classical domain-wall model, Bhattacharjee\cite{Bhattacharjee}
showed that dislocations can induce a crossover and change the
nature of the C-IC transition. This problem will not be discussed
further in this paper.

\subsection{Comparison with experiments}
The present argument on quantum melting of stripes are rather
general and hence can be applied to various 2D systems that have
incommensurate stripe structures, e.g., adsorbed atoms on graphite
and doped high-T$_{\rm c}$ cuprates. Comparing with experiments
done for various adsorbed atoms, we realize that quantum melting
of stripes seems to have been already observed in scattering
experiments of H$_2$/Gr and D$_2$/Gr,\cite{Freimuth} where the
correlation length shows a sudden decrease in a finite density
region just above the commensurate density, though this region has
not been regarded as the stripe liquid phase previously.

We can also find a signal for melting of the stripe structure in
scattering experiments\cite{Yamada,Fujita} on
La$_{2-x}$Sr$_x$CuO$_4$, where the width of incommensurate peaks
becomes larger with lowering the hole density in $x < 1/8$. To
understand the Sr-concentration dependence of the peak width, we
might need to take into account the strength and form of
interactions more carefully. The case of power-decay interactions
is discussed in Sec.~\ref{sec:power}. It should be noted that
effects of randomness are important in some experiments. Bogner
and Scheidl\cite{Bogner} discussed that randomness can also make
stripes short-range ordered for any wall spacing.

\section{Honeycomb phase}\label{sec:honeycomb}
In this section we discuss quantum fluctuations of the honeycomb
structure. First, we consider local lattice vibrations. Since each
straight line has finite length $l$ in hexagons, the zero-point
energy of one domain wall has a finite-size correction
\begin{equation}\label{ZPE_line}
\Delta \zeta_0(l)= - \dfrac{\pi \hbar}{16}
\sqrt{\dfrac{\gamma}{m}} \dfrac{1}{l^2}.
\end{equation}
There are also so-called ``breathing modes'',\cite{Villain} in
which the classical energy except interactions does not change
with enlarging one hexagon. In quantum mechanics, this degree of
freedom creates a zero-point oscillation and hence increases the
energy. If the mass $m$ is large and the deviation of $u$ is very
small, the potential from the adjoining parallel walls can be well
approximated by a quadratic form and in the absence of meandering
of walls, breathing of one hexagon behaves as a harmonic
oscillator. The zero-point energy per cell is thus roughly
estimated in the exponential decay form $\exp ( -\sqrt{3}\kappa
l/2 )$ for large $m$. On the other hand, the hard-core repulsion
is dominant for small $m$. In the harmonic approximation the
problem is reduced to a quantum particle in the square well
infinite potential and the energy decays as $1/(m l^3)$ per cell.

Global lattice vibrations of the honeycomb structure can be
treated with a continuum elastic theory, in which the displacement
vector ${\bf u}=(u_x,u_y)$ of domain walls (or intersections)
obeys the Hamiltonian
\begin{eqnarray}
H&=&\dfrac{1}{2}\int d^2 r \Biggl\{ \dfrac{\sqrt{3} l}{2m} {\bf p
}^2 + \dfrac{\mu}{2}(\partial_i u_j +\partial_j u_i)^2
+ \lambda ({\bf \nabla} \cdot {\bf u} )^2 \nonumber\\
& & + \gamma' ({\bf \nabla} \times {\bf u})^2 \Biggr\}.
\label{eq:elas_h}
\end{eqnarray}
UV cutoffs of the vibrations are set as $\Delta x=\Delta y=Ql$.
The time cutoff for the longitudinal mode is given by $\Delta
\tau=a/c_{\rm l}$ with ${c_{\rm l}}^2=\sqrt3 l(\mu+\lambda/2)/m$,
and for the transverse mode $\Delta \tau=a/c_{\rm t}$ with
${c_{\rm t}}^2=\sqrt3 l(\mu+\gamma')/2m$. The Lam\'e coefficient
$\gamma'$ can be estimated as $\gamma'= 2\gamma/\sqrt{3}l$ from
the anisotropy of the substrate, where $\gamma$ is the same
interfacial stiffness as in stripes. Other coefficients $\mu$ and
$\lambda$ are determined from the similar procedure as that for
the stripe structure.

In the large-$m$ limit, the Lam\'e coefficients $\mu$ and
$\lambda$ are determined by the exponential repulsion in the form
\begin{equation}
2\mu+\lambda \simeq (54 \alpha \kappa^3 /\pi^2)l^3 \exp(-\sqrt{3}
\kappa l).
\end{equation}
For small $m$, the hard-core repulsion between walls (and
intersections) creates a steric force. To estimate Lam\'e
coefficients $\mu$ and $\lambda$ due to this steric force, we
extend Helfrich's idea\cite{Helfrich} to this problem: We made a
self-consistent condition
\begin{equation}\label{self-consist}
l^2 \dfrac{\partial^2 \Delta e(\mu,\lambda)}{\partial l^2} =
4(\mu+\lambda),
\end{equation}
where $\Delta
e(\mu,\lambda)=e(\mu,\lambda,\gamma')-e(0,0,\gamma')$ and
$e(\mu,\lambda,\gamma')$ denotes the energy of eq.\
(\ref{eq:elas_h}) given by
\begin{eqnarray}
e(\mu,\lambda,\gamma') &=& \dfrac{\hbar \pi^2}{12 l^3}
\sqrt{\dfrac{\sqrt{3}\mu l}{2m} + \dfrac{\gamma}{m}}
\nonumber\\
&+& \dfrac{\hbar \pi^2}{12 l^3}\sqrt{\dfrac{\sqrt{3}(2\mu +
\lambda) l}{2m}}. \label{eq:zpe_h}
\end{eqnarray}
Under the assumption $\mu = b \lambda$, eq.\ (\ref{self-consist})
is solved in the form
\begin{equation}\label{energy_h}
\mu = b \lambda \approx \biggl(\dfrac{35 \pi^5}{12}\biggr)^2
\dfrac{\sqrt{3} b (b+1/2)}{(b+1)^2} \dfrac{\hbar^2}{m l^5},
\end{equation}
where the ratio $b$ cannot be determined. This form implies that
the wall-wall (or intersection-intersection) interaction
effectively decays in the power decay form $1/l^3$. From these
Lam\'e coefficients, and using eqs.\ (\ref{ZPE_line}) and
(\ref{eq:zpe_h}), the zero-point energy is estimated. For large
$l$, the leading order term of the energy from the global modes is
more dominant than the one that comes from the breathing mode for
any $m$. The breathing oscillation, thus, gives a soft mode at
$T=0$. The most dominant term of the zero-point energy is
\begin{equation}
\Delta E_{\rm h} \simeq \dfrac{2 \Delta \zeta_0(l)}{\sqrt{3}l}
 + \dfrac{\hbar \pi^2}{12 l^3} \sqrt{\dfrac{\gamma}{m}}
\end{equation}
for any $m~(<\infty)$, i.e. regardless of forms of the dominant
interaction. The second term can be considered as the zero-point
energy of torsion vibrations of hexagons.

This system has two components ($u_x,u_y$) in $xy$ plane and hence
has similar degrees of freedom as vortex melting problems in 3D
superconductors at finite temperature, in which it has been argued
that the transition is of first order.\cite{BrezinNT,Blatter} We,
hence, expect that the quantum melting of the honeycomb structure
at $T=0$ is also of first order and one cannot reach a critical
region. Nevertheless it is instructive to derive the phase
boundary assuming a second-order transition. In the 2D melting
problem, the leading-order term of the bare
stiffness\cite{NelsonH,Coppersmith} is given by $(Ql)^2 \mu [
(\mu+\lambda)/(2\mu+\lambda) + 1 ]$ and that becomes $(Ql)^2\mu$
in the strong anisotropy limit whatever $b$ is. The most dominant
stiffness, thus, comes from $\mu$ in the 2D plain. In the 2+1D
system, after the rescaling $x'=ax/Ql$, $y'=ay/Ql$, ${\bf u}'=2\pi
{\bf u}/Ql$, and $\tau'=c_{\rm t} a \tau/Ql$ for the transverse
mode and $\tau'=c_{\rm l} a \tau/Ql$ for the longitudinal one, the
effective action has the bare stiffness
\begin{equation}
\tilde{K}_{\rm l}= Q^2 l m c_{\rm l}/\sqrt{3}(2\pi)^2
\end{equation}
for the longitudinal mode, and
\begin{equation}
\tilde{K}_{\rm t}=Q^2 l m c_{\rm t}/\sqrt{3}(2\pi)^2
\end{equation}
for the transverse one. We expect that, even in the 2+1D system,
the most dominant contribution to the stiffness in $xy$ plane
comes from $\mu$ and the effective coupling constant is presumably
given by
\begin{equation}
\tilde{K}\approx Q^2 l \sqrt{l m \mu}/3^{1/4} (2\pi)^2 \sim \hbar
Q^2/l
\end{equation}
in the strong anisotropy limit ($\mu,\lambda\ll\gamma'$). This
indicates that the honeycomb structure is unstable for
$\tilde{K}/\hbar < C$ with a constant $C$ of order unity, and
hence it melts for large $l$ region.

\section{Phase Diagrams for Adsorbed Atoms}\label{sec:PD}
From the above arguments, we obtain phase diagrams for
incommensurate phases of adsorbed atoms.\cite{Momoi} A phase
diagram with the parameters $m^{-1}$ and $\delta n$ was shown in
Fig.\ \ref{fig:pd_final}(a) for a positive but not very large
intersection energy $f_{\rm I}$. Various phase diagrams can appear
depending on the intersection energy $f_{\rm I}$, which is
determined by microscopic atomic configurations around
intersections.

The stripe liquid phase exists for any mass in the vicinity of the
onset of the C-IC transition and no direct transition can occur
from the commensurate phase to the stripe ordered one. Actually
this stripe liquid phase was presumably observed in H$_2$ and
D$_2$ on graphite,\cite{Freimuth} as already mentioned in Sec.\
\ref{sec:stripe}D. The phase boundary between the stripe liquid
and stripe (ordered) solid phases shows a second-order transition
in the 3D XY universality class. The phase boundary shows a
reentrant behavior because of the crossover of dominant
interactions around $\omega\approx 2$ as discussed in Sec\
\ref{sec:stripe}C. The crossover mass can be roughly estimated
from $\omega_{\rm c}=2$. For this purpose, we use the value of the
domain-wall width $\kappa^{-1}$ estimated by Villain\cite{Villain}
as $\kappa^{-1}=5a$ for ${}^3$He monolayer on graphite. The
stiffness $\gamma$ comes from the domain-wall energy and it should
be of the same order as the melting temperature of the
commensurate phase. We hence set $a\gamma\approx\mbox{a few [K]}$
and then roughly estimate that the cross-over mass ($am_{\rm
c}\approx a\hbar^2 \kappa^4 /4(4\pi)^2 \gamma$) is about $10^{-4}$
[atomic mass unit] in monolayer on graphite. We note that this
value can vary greatly by changing the substrate potential.

If the intersection energy $f_{\rm I}$ is not very large, this
melting line merges into the phase boundary between the stripe and
honeycomb phases. Because the transition from the stripe phase to
the honeycomb one is of first order, the system shows a phase
separation, in which two phases coexist, for a certain density
region (the shaded ones in Fig.\ \ref{fig:pd_final}). From
comparison of the zero-point energies of two phases, this phase
boundary slightly moves to higher density region with lightening
the mass. Moreover, in the small-mass limit, commensurate domains
themselves would melt and become a conventional isotropic liquid.
This isotropic liquid phase is also taken into account in Fig.\
\ref{fig:pd_final}.

Another quantum correction might appear in the width $\kappa^{-1}$
of domain walls. Halpin-Healy and Karder\cite{Halpin-Healy} argued
that, in light atoms, the commensurate solid is more compressible
due to quantum fluctuations and the width $\kappa^{-1}$ becomes
small, whereas, in heavy atoms, the width $\kappa^{-1}$ is large.
This is consistent with the estimates from experiments, where the
width was estimated as $\kappa^{-1}\approx 0.8 a$ for D$_2$
systems\cite{Freimuth} and $\approx 5.7a$ for Kr
systems.\cite{Stephens} In real adsorbed systems, it is hence
likely that due to $m$-dependence in $\kappa^{-1}$ the honeycomb
phase is favored for heavy atoms and the stripe phase for light
atoms [see Fig.\ \ref{fig:pd_final}(b)].

\section{Stripes with long-range interactions}\label{sec:power}
In this section we consider the case that stripes are repelling
with each other by long-range interactions with a power-decay
form. For example, in doped antiferromagnets, if holes are
perfectly doped only in domain walls and domains are insulating,
there must be Coulomb long-range repulsion between hole-rich
domain walls.

Let us consider the power-decay interaction $V(u)=c_1 u^{-q+1}$
between walls, where $c_1$ is a positive constant. (Coulomb
repulsion between lines corresponds to $q=1$, i.e. $V(u)=-\ln u$.)
In the case of Coulomb repulsion, however, the plasma modes
$\omega_{\rm p}\sim \sqrt{k_x}$ appear and low-lying excitations
in $k_x$-direction does not have $k$-linear spectrum. The plasma
modes appear for the long-range interactions $r^{-q+1}$ with $q\le
2$ (see Appendix B). Hence for stripes with $q\le 2$, the
stiffness $K_x$ diverges and the elastic theory with the
Hamiltonian (\ref{eq:H}) cannot be applied. Hereafter, we consider
only the case $q>2$ and use the melting criterion derived in sec.\
\ref{sec:stripe}B. The renormalization-group
equation\cite{FisherH} shows that power-law repulsions are not
modified by the entoropical force from the hard-core repulsion.
The stiffness $K_x$ of the Hamiltonian (\ref{eq:H}) is hence given
by
\begin{equation}
K_x = c_2 l^{-q}
\end{equation}
with $c_2>0$ for any parameter $\omega$. After the rescaling
$\tilde{u}=2\pi u/Ql$, $\tilde{\tau}=c\tau$, and
$\tilde{x}=ax/Ql$, we have the effective coupling
\begin{equation}\label{eq:stiffness_H}
\tilde{K}_x = \dfrac{Qa c_2}{4\pi^2 c l^{q-1}}
\end{equation}
for the effective action (\ref{rescaled-action}).

\begin{figure}[tbp]
\includegraphics[width=2.5in]{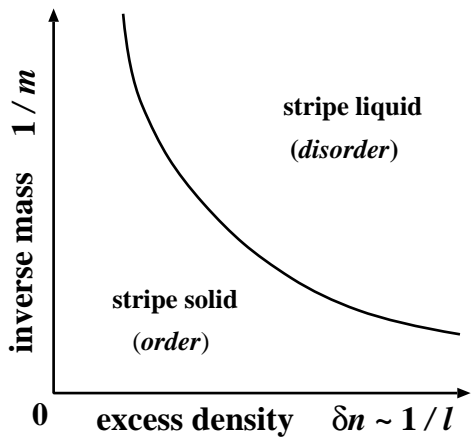}
\caption{\label{fig:pd_stripeLR}Phase diagram of the stripe
structure with long-range interaction $r^{-q}$ with $2<q<5$.}
\end{figure}

Inserting eq.\ (\ref{eq:stiffness_H}) into eq.\
(\ref{eq:criterion}), we find that the critical point $l_{\rm c}$
satisfies
\begin{eqnarray}
\lefteqn{\left.(\tilde{K}_x \tilde{K}_y
\tilde{K}_\tau)^{1/3}\right|_{l=l_c}\hbar^{-1}} \nonumber\\
 &=& \dfrac{1}{4\pi^2 \hbar} \left\{ \dfrac{Q^7 l_c^{5-q}
\gamma^{1/2} m^{3/2} c_2}{a} \right\}^{1/3}=C,
\end{eqnarray}
where $C$ is a positive constant of order unity. From the melting
criterion, we thus find that stripes with $2<q<5$ melt for
small-wall-spacing (high-density) region and stabilize for
large-wall-spacing (low-density) region because of long-range
repulsion. Note that this behavior is contrary to the short-range
case. A phase diagram is shown in Fig.\ \ref{fig:pd_stripeLR}.
This tendency to solidification in low-density region is similar
to that of charged point objects, e.g. Wigner crystals, and may be
universal in systems with Coulomb repulsion. On the other hand,
for $q>5$ the tendency to striped ordering is the same as the
short-range interaction system and stripes become a liquid for
low-density region.

\section{Discussions}\label{sec:discuss}
To summarize, we studied the effect of quantum fluctuations on 2D
domain walls and showed a quantum-melting behavior. We found that
striped domain walls with short-range interactions become a stripe
liquid even at zero temperature in the incommensurate phase in the
vicinity of the onset of C-IC transition. We discussed both
short-range and long-range interactions that decay in power laws.
It was revealed that striped correlation is enhanced in high
density for the case of short-range interactions and in low
density for the case of long-range interactions. Though we
employed various approximations in estimating the stiffness $K_x$,
we believe that the above qualitative conclusion does not
drastically depend on the choice of approximations.

We discussed mass dependence of domain-wall structures in adsorbed
atoms and obtained phase diagrams as shown in Fig.\
\ref{fig:pd_final}. For this purpose, we assumed that the
intersection energy is positive. Recently Morishita and
Takagi\cite{MorishitaT2} performed path-integral Monte Carlo
simulations for $^3$He on graphite. Their result is consistent
with the positive intersection energy. They also proposed a
honeycomb-cage structure where commensurate regions and
discommensurate ones are opposite to the honeycomb incommensurate
structure.\cite{MorishitaT} Direct STM measurements of ${}^3$He
atoms in the incommensurate phase are desired to clarify the real
structure.\cite{Fukuyama} As we commented in Sec.~\ref{sec:PD},
the quantum correction to the domain-wall width $\kappa^{-1}$ may
be also important to reproduce precise phase diagrams of
incommensurate adsorbed atoms on graphite. Evaluation of $m$
dependence of $\kappa$ remains to be considered.

In a doped Mott insulator, Kivelson {\it et al.}\cite{KivelsonFE}
discussed appearance of electronic nematic and smectic phases from
an analogy of classical liquid crystals. From the symmetry, the
stripe liquid we discussed is characterized as a nematic state.
Our arguments thus demonstrated appearance of a quantum nematic
state at $T=0$ for low-density region and the correlation length
to be finite until zero temperature even in pure systems. A
quantum transition from the electronic nematic phase to isotropic
Fermi liquid was also discussed by Oganesyan {\it et
al}.\cite{OganesyanKF}

Finally, we comment implication of our results on the doping
dependence of the incommensurate order observed in
La$_{2-x}$Sr$_x$CuO$_4$.\cite{Yamada,Fujita} Since
incommensurability is proportional to Sr concentration $x$ in the
region $x<1/8$, our domain-wall model can be applied in this
region. The system is in the insulating spin-glass phase for
$x<0.055$ and the superconducting phase for $x>0.055$. The
correlation length gradually decreases with increasing
Sr-concentration $x$ from $x\approx 0.024$ and after the system
enters into the superconducting phase the correlation length turns
to increase. If we assume that antiferromagnetic domains are
insulating in the spin-glass phase and metallic in the
superconducting one (above $T_{\rm c}$), the wall-wall interaction
would show the power-decay and exponential ones, respectively.
Under this assumption, the hole-density dependence of the
correlation length can be naturally understood from the present
results. More precisely, in Coulomb repulsive systems the
stiffness $K_x$ is infinite in pure systems due to plasma modes
and randomness would be relevant to make the correlation length
finite\cite{Bogner}. Another possibility would be that,
antiferromagnetic domains are always insulating for $x<1/8$ and
there is power-decay force between stripes. In that case, to
understand enhancement of the correlation length in $0.055<x<1/8$,
we may need to consider that the wall-wall interaction is enhanced
in the superconducting phase due to coupling with fluctuations of
superconductivity. This is a current hot topic, but out of scope
of the present paper.

\begin{acknowledgments}
The author would like to thank Hiroshi Fukuyama, D.\ R.\ Nelson,
E.\ Demler, Y.\ Ohashi, Y.\ Nishiyama, H.\ Yoshino, M.\ Morishita
and M.\ Fujita for stimulating discussions. He also acknowledges
kind hospitality of the condensed matter theory group in Harvard
University, where this research was partially done. This work was
supported in part by Monbusho (MEXT) in Japan through Grand Nos.\
13740201 and 1540362.
\end{acknowledgments}

\appendix

\section{Renormalization-group treatment of a string between  flat walls}
\label{RG_one_string} Using the functional renormalization-group
treatment given by Fisher and Huse,\cite{FisherH} we reexamine the
calculation of effective potential between a quantum string and a
flat wall given in Ref.\ \onlinecite{Sornette}. For completeness,
we briefly show the calculation. Instead of using the external
potential used in Ref.\ \onlinecite{Sornette}, we consider a
valley potential
\begin{equation}
V(x)=\left\{ \begin{array}{lc}
  A,                 & \mbox{\hspace*{1cm}}|x|>l, \\
  B \exp\{-\kappa (x+l)\} \nonumber\\
  ~~~+ B \exp\{-\kappa (l-x)\}, &
  \mbox{\hspace*{1cm}}|x|<l,
\end{array} \right.
\end{equation}
where the spacing $l$ between the string and walls is fixed in
advance. Rescaling the imaginary time $\tau$ as $c\tau$ with
$c=\sqrt{\gamma/m}$, we have the isotropic effective action for
one quantum string in the potential
\begin{equation} S=\dfrac{\sqrt{m \gamma}}{2\hbar}\int d\tau dy
\{ (\partial_\tau u)^2 + (\partial_y u)^2 + \gamma^{-1} V(u) \}.
\end{equation}
Integrating out fast modes in string fluctuation up to UV cutoff
$\Delta \tau = \Delta y = \xi$, we obtain the renormalized action
\begin{equation}
S^{\rm R}=\dfrac{\sqrt{m \gamma}}{2\hbar}\int dt dy \{
(\partial_\tau u)^2 + (\partial_y u)^2 + \gamma^{-1} V_\xi (u) \}
\end{equation}
with the renormalized potential\cite{FisherH}
\begin{equation}
V_\xi (u)=\dfrac{1}{\sqrt{2\pi}\delta(l^*)} \int^\infty_{-\infty}
du' V(u') \exp \left\{-\dfrac{(u-u')^2}{2\delta^2 (l^*)}\right\},
\end{equation}
where $\delta(l^*)=\hbar l^*/\pi\sqrt{m\gamma}$ and $l^*=\log
(\xi/a)$. The correlation length is determined by\cite{BrezinHL}
\begin{equation}\label{eq:cond_xi}
\left.\dfrac{\partial^2 V_\xi (u)}{\partial u^2}
\right|_{u=0}=\dfrac{\gamma}{\xi^2}.
\end{equation}

One can evaluate the renormalized potential using the
steepest-decent method in the large-$l$ limit. For
$\omega~(=\hbar\kappa^2/4\pi \sqrt{m\gamma})>2$,
\begin{widetext}
\begin{eqnarray}
V_\xi (u)&=&\sqrt{\dfrac{l^* \omega}{\pi}} \left\{
\dfrac{A}{\kappa(l+u)} + \dfrac{B}{2l^*\omega-\kappa(l+u)}
\right\}
\exp\left\{ -\dfrac{\kappa^2}{4l^* \omega} (l+u)^2\right\}\nonumber\\
&+& \sqrt{\dfrac{l^* \omega}{\pi}} \left\{ \dfrac{A}{\kappa(l-u)}
+ \dfrac{B}{2l^*\omega-\kappa (l-u)} \right\} \exp\left\{
-\dfrac{\kappa^2}{4l^*\omega } (l-u)^2\right\}.
\end{eqnarray}
\end{widetext}
The condition (\ref{eq:cond_xi}) gives the correlation length in
the leading order
\begin{equation}\label{eq:corr1} \xi = \sqrt{a}
\Biggl(\dfrac{\gamma^2 \omega^{3/2}l}{\kappa^3}\Biggr)^{1/8} \exp
\Biggl(\dfrac{\kappa l}{2\sqrt{2\omega}}\Biggr)
\end{equation}
and the renormalized potential $V_\xi$ at the equilibrium position
$u=0$
\begin{equation}\label{eq:poten1}
V_\xi (u)\biggr|_{u=0} = \dfrac{A}{a \sqrt{\pi}}
\Biggl(\dfrac{2\kappa^3\gamma^2}{\omega^{3/2} l}\Biggr)^{1/4} \exp
\Biggl(-\dfrac{\kappa l}{\sqrt{2\omega}}\Biggr),
\end{equation}
where the exponents of the power forms in eqs.\ (\ref{eq:corr1})
and (\ref{eq:poten1}) are different from those in
Ref.~\onlinecite{Sornette}.

For $\omega<2$, $V_\xi (u)$ is calculated as
\begin{eqnarray}
V_\xi (u)&=&\dfrac{\sqrt{l^*\omega} A}{\sqrt{\pi}\kappa(l+u)}
\exp\left\{ -\dfrac{\kappa^2}{4l^*\omega } (l+u)^2\right\}\nonumber\\
&+& \dfrac{\sqrt{l^*\omega} A}{\sqrt{\pi}\kappa(l-u)}
\exp\left\{ -\dfrac{\kappa^2}{4l^*\omega} (l-u)^2\right\}\nonumber\\
&+& B \exp \left\{ -\kappa(l+u) + l^*\omega \right\}\nonumber\\
&+& B \exp \left\{ -\kappa(l-u) + l^*\omega \right\}.
\end{eqnarray}
The condition (\ref{eq:cond_xi}) leads to the correlation length
\begin{equation}
\xi=a\left( \dfrac{\gamma}{2B\kappa^2 a^2} \right)^{1/(2+\omega)}
\exp \left( \dfrac{\kappa l}{2+\omega} \right)
\end{equation}
and the renormolized potential at the equilibrium position $u=0$
\begin{equation}
V_\xi (u)\biggr|_{u=0} = 2B \left( \dfrac{\gamma}{2B\kappa^2 a^2}
\right)^{\omega/(2+\omega)} \exp \left(-\dfrac{2\kappa l}
{2+\omega} \right).
\end{equation}
These are the same as the previous ones.\cite{Sornette}

\section{Elastic theory of stripes with long-range interactions}
If we have a stripe structure of walls displaced slightly from
equilibrium, then the additional potential energy is
\begin{equation}
 \Delta E= \dfrac{1}{2} \sum_{{\bf r}_i\ne {\bf r}_j} |{\bf r}_i-{\bf r}_j|^{-q},
\end{equation}
where
\begin{equation}
\sum_{{\bf r}_i} \equiv \sum_i \int dy_i
\end{equation}
and
\begin{equation}
 {\bf r}_i = (il+u_i(y_i),y_i).
\end{equation}
Here $u_i(y_i)$ denotes displacement at $y=y_i$ in $i$th wall. If
the displacements are small, we can expand the energy in powers of
$u$ and obtain\cite{BonsallM,FisherHM}
\begin{equation}\label{eq:elastic}
 \Delta E= \dfrac{1}{2} \sum_{{\bf r}_i, {\bf r}_j} \Pi({\bf r}_i,{\bf r}_j)
 u({\bf r}_i)u({\bf r}_j),
\end{equation}
where $u({\bf r}_i)=u_i(y_i)$. We can Fourier transform
(\ref{eq:elastic}) to obtain
\begin{equation}
 \Delta E= \dfrac{1}{2(la)^2} \int \dfrac{d^2 p}{(2\pi)^2} \Pi({\bf p})
 |u({\bf p})|^2
\end{equation}
and
\begin{eqnarray}
 \Pi({\bf p})&=&la \lim_{z\to 0} \dfrac{\partial^2}{\partial z^2} \sum_{\bf r}
 \dfrac{1-\exp(-i p_x r_x)}{|z {\bf e}_x - {\bf r}|^q}\\
 &=& \left\{
  \begin{array}{cc}
    \dfrac{2 \pi^{3/2}\Gamma( (q-1)/2 )}{(q-2)\Gamma( q/2
 )^2 l^{q-2}}p_x^2, & ~~~q>2, \\
    c_1 p_x^2\log p_x, & ~~~q=2\\
    c_2 |p_x|, & ~~~q=1
  \end{array}
\right.
\end{eqnarray}
for small $p_x$, where $c_1$ and $c_2$ are positive constants. On
the other hand, in the elastic theory, potential energy is written
as
\begin{equation}
 \Delta E= \dfrac{K_x}{2} \int \dfrac{d^2 p}{(2\pi)^2} |p_x|^2
 |u({\bf p})|^2.
\end{equation}
Thus if the exponent satisfies $q>2$, we can describe excitations
with the elastic theory of Gaussian forms and, if $q\le 2$, we
cannot because $K_x$ diverges.



\begin{thebibliography}{99}
\bibitem{MF} Here, ``mean-field approximation'' means that two
adjoining walls are set to straight and only fluctuations of a
single wall between them are treated.
\end{thebibliography}
\end{document}